\documentclass[preprint,11pt]{article}
\pdfoutput=1
\usepackage{authblk}

\usepackage{amsmath,latexsym,amssymb,color,epsfig}
\usepackage{dcolumn}% Align table columns on decimal point
\usepackage{bm}
\usepackage{cancel}
\usepackage{caption}
\usepackage[english]{babel}
\usepackage{array}
\usepackage[ansinew]{inputenc}
\usepackage{textcomp}
\usepackage{footnote}

\usepackage[norule]{footmisc}

\usepackage[sorting=none]{biblatex} %Imports biblatex package
\newcommand{\be}{\begin{equation}}
\newcommand{\ee}{\end{equation}}
\newcommand{\bea}{\begin{eqnarray}}
\newcommand{\ena}{\end{eqnarray}}

\newcommand{\nb}{\nonumber}

\newcommand{\de}{\partial}

\newcommand{\ba}{\begin{eqnarray}}
\newcommand{\ea}{\end{eqnarray}}

\makeatletter
\def\ps@mine{%
    \def\@oddfoot{\hfil\thepage\hfil}\let\@evenfoot\@oddfoot
    \let\@oddhead\@evenhead%
    \let\@mkboth\@gobbletwo
    \let\sectionmark\@gobble
    \let\subsectionmark\@gobble
  }

\renewcommand\section{\@startsection {section}{1}{\z@}%
                                   {-3.5ex \@plus -1ex \@minus -.2ex}%
                                   {2ex \@plus.2ex}%
                                   {\normalfont\large\sffamily\bfseries}}
\renewcommand\subsection{\@startsection {subsection}{1}{\z@}%
                                   {-3.5ex \@plus -1ex \@minus -.2ex}%
                                   {2ex \@plus.2ex}%
                                   {\normalfont\sffamily\bfseries}}

\makeatother
\numberwithin{equation}{section}

\makeindex
\begin{document}
\hyphenrules{nohyphenation}

\thispagestyle{empty}
\vspace*{-2.5cm}
\begin{minipage}{.45\linewidth}

\end{minipage}
\vspace{2.5cm}

\begin{center}
{\huge\sffamily\bfseries The Physical Content of Long Tensor Modes\\ in Cosmology}
\end{center}

 \vspace{0.5cm}
 
 \begin{center}
 {\sffamily\bfseries \large Nicola Bartolo}$^{a,b,c}$,
 {\sffamily\bfseries \large Giovanni Battista Carollo}$^{a,d}$, 
 {\sffamily\bfseries \large Sabino Matarrese}$^{a,b,c,e}$,  
 {\sffamily\bfseries \large\\  Luigi Pilo}$^{f,g}$,
  {\sffamily\bfseries \large Rocco Rollo$^{f, h}$}\\[2ex]
  {\it $^a$ Dipartimento di Fisica e Astronomia ``G. Galilei",
Universit\`{a} degli Studi di Padova, via Marzolo 8, I-35131 Padova,
Italy\\\vspace{0.1cm}
$^b$ INFN, Sezione di Padova, via Marzolo 8, I-35131 Padova, Italy\\\vspace{0.1cm}
$^c$ INAF-Osservatorio Astronomico di Padova, Vicolo dell'Osservatorio 5, I-35122 Padova, Italy\\\vspace{0.1cm}
$^d$ Dipartimento di Fisica ``M. Merlin", Universit\`{a} degli Studi di Bari, Via Giovanni Amendola, 173, 70125 Bari, Italy\\\vspace{0.1cm}
$^e$ GSSI-Gran Sasso Science Institute, Viale Francesco Crispi, 7, 67100 L'Aquila, Italy\\\vspace{0.1cm}
$^f$ Dipartimento di Scienze Fisiche e Chimiche, Universit\`a degli Studi dell'Aquila,  I-67010 L'Aquila, Italy\\\vspace{0.1cm}
$^g$ INFN, Laboratori Nazionali del Gran Sasso, I-67010 Assergi, Italy\\\vspace{0.3cm}
$^h$ Centro Nazionale INFN di Studi Avanzati GGI, Largo Enrico Fermi 2,  I-50125 Firenze, Italy\\\vspace{0.1cm}
{\tt nicola.bartolo@pd.infn.it},
{\tt giovannibattista.carollo@studenti.unipd.it}}, {\tt luigi.pilo@aquila.infn.it}, {\tt sabino.matarrese@pd.infn.it}, {\tt rocco.rollo@gssi.it
}
\end{center}

\vspace{0.7cm}

\begin{center}
{\small \today}
\end{center}

\vspace{0.7cm}

\begin{center}
{\bf \Large Abstract}\\
\end{center}
We analyze the physical content of squeezed bispectra involving long-wavelength tensor perturbations, showing that these modes cannot be gauged away, except for the exact (unphysical) limit of infinite wavelength, $k=0$. This result has a direct implication on the validity of the Maldacena consistency relation, respected by a subclass of inflationary models. Consequently, in the squeezed limit, as in the case of the scalar-scalar-scalar bispectrum, squeezed mixed correlators could be observed by future experiments, remaining a key channel to study Early Universe physics and discriminate among different models of inflation.

\newpage

\section{Introduction}
The study of primordial non-Gaussianity (PNG) is one of the  most important avenues to probe inflation, trying to resolve the large degeneracy among different models still present after analyzing Cosmic Microwave Background (CMB) data. As well-known, the amount of non-Gaussianity in standard single-field slow-roll inflation is very tiny, being of the order of the slow-roll parameters (\cite{Gangui:1993tt,Gangui:1999vg,Wang:1999vf,Acquaviva:2002ud,Maldacena:2002vr,Lyth:2005du}), yet non-vanishing; on the other hand, a large class of multi-field theories leads to quite different predictions (\cite{Seery:2005gb,Gao_2008,Byrnes_2010,Garcia_Saenz_2020}), as well more general single-field models of inflation~\cite{Seery_2005,Chen:2006nt,Senatore:2009gt,planckcollaboration2019planck}. \\
The strength of non-Gaussianity $f_{\rm NL}$ is the key parameter to quantify the phenomenon (\cite{planckcollaboration2019planck}), being related to the bispectrum, which vanishes for a perfectly Gaussian field. 
In the case of single-field ``standard inflation", $f_{\rm NL}$ contains in particular a {\it local} contribution, which is maximum for {\it squeezed} bispectrum triangles, where one wave-number is much shorter than the other two.
In this context, one of the main results is the so-called {\it Maldacena consistency relation} (\cite{Maldacena:2002vr,Creminelli_2004,Creminelli:2004pv,Cheung:2007sv,Creminelli:2011rh,Bartolo:2011wb,Senatore:2012wy}), stating that in the squeezed limit the bispectrum (for {\it any} single-field model of inflation) becomes simply a product of two power-spectra
\begin{equation}
\begin{aligned}
\lim _{k_1\rightarrow 0}\langle\zeta(\vec {k}_{1}) \zeta(\vec {k}_{2}) \zeta(\vec {k}_{3})\rangle =- (2 \pi)^{3} \delta^{(3)}(\vec{k}_{1}+\vec{k}_{2}+\vec{k}_{3})   (n_s-1)  P_\zeta(k_1)P_\zeta(k_2) \ ,
\end{aligned}
\label{consistencyrelation}
\end{equation}
where $\zeta$ is the comoving curvature perturbation\footnote{In our convention, $\zeta$  is defined from the Ricci scalar curvature $\mathcal R_S$ of an hypersurface of equation $S(x)=$const. where $S$ is a four-dimensional scalar, according with
$$
\mathcal R_S=-\frac{4}{a^2}\nabla^2 \zeta \ ,
$$
at first order in perturbations (where $a$ is the scale factor defined in (\ref{FLRW})).\label{footnote1}}, $P_\zeta$ its power spectrum and $n_s$ the scalar spectral index. This consistency relation has been derived, using different approaches, such as path-integration (\cite{Goldberger_2013}), exploiting the residual symmetries of the gauge-fixed action for $\zeta$ (\cite{Creminelli:2012ed,Hinterbichler:2012nm,Hinterbichler:2013dpa,Hui:2018cag}), BRST symmetry (\cite{Binosi,Berezhiani_2014}) and holography (\cite{Schalm,Bzowski_2013}).\\
A similar consistency condition is valid for any type of bispectra in (single-field) inflation, including both the curvature perturbation and the tensor modes (\cite{Maldacena:2002vr,Hinterbichler:2013dpa,Bordin:2016ruc}). There are however models for which the consistency relation is violated. For example, it has been explicitly shown that some inflationary models involving more than one scalar field (\cite{GordonWands}), a non-attractor phase (\cite{Lindefast,Hossein,Kinney}), an unstable background (\cite{Brahma,KhouryPiazza}) or breaking of space-time diffeomorphism invariance (\cite{SolidInflation,Bartolo:2015qvr,Celoria_2021, celoria2021primordial}) violate the consistency condition. This implies that, from the phenomenological point of view, the consistency relation is a very interesting channel to study Early Universe physics, given that it links observable quantities: any deviation from it would rule out all single-field models of inflation and could imply that one of the previous assumptions is valid. The {\it Planck} analysis of the CMB temperature and E-mode polarization provided, among the various results, the following limit on the non-Gaussianity strength for the local shape (\cite{planckcollaboration2019planck}) of curvature bispectrum: $f^\text{local}_\text{NL}=-0.9 \pm  5.1$ at $68 \%$ C.L.  When compared to the spectral index, $n_s = 0.9652 \pm 0.0042$ ($68 \%$ C.L.), it is clear that the consistency relation is far from being tested from an experimental point of view.\\
In the last decade a debate has emerged (\cite{Tanaka_2011, Pajer:2013ana, Dai:2015jaa, Dai:2015rda, Bravo:2017gct}) on the observability of squeezed bispectra: various groups have claimed that the consistency relations can be gauged away by a suitable rescaling of the spatial coordinates and, as a result, they cannot be considered as physical observables. In particular, the key-ingredient to cancel squeezed bispectra is the passage to the so-called Conformal Fermi Coordinates (CFC) frame (\cite{Pajer:2013ana,Dai:2015rda}). The very same technique was used to cancel any squeezed $\zeta$-related quantity and, as a consequence, also the halo bias scale-dependence, as far as the so-called ``GR-contribution" (see~\cite{Bartolo:2005xa}) is concerned (\cite{Dai:2015jaa,Baldauf_2011,dePutter:2015vga,Cabass:2018roz}). 
%In the same way, if the tensor bispectra are unphysical, this would have %relevance for correlating the CMB temperature to the stochastic %gravitational-wave background (\cite{Malhotra:2020ket, %Adshead:2020bji}). 
Moreover, tensor fossils (\cite{Giddings:2010nc,Masui:2010cz,Giddings:2011zd,Jeong:2012df,Dai:2013ikl,Dai:2013kra,Dimastrogiovanni:2014ina, Dimastrogiovanni_2016,Dimastrogiovanni:2019bfl}) in single-field inflationary models have been claimed to be not genuine physical quantities~\cite{Pajer:2013ana,Brahma}.\\
In this paper we argue that the gauge freedom used to cancel squeezed correlation functions is only valid if the squeezed momentum is {\it exactly zero}. As shown in \cite{Matarrese:2020why}, when the squeezed momentum is finite, the gradient expansion restores the consistency relations. In \cite{Matarrese:2020why} the analysis was limited only to the scalar sector, but here we argue that the same result applies to the tensor sector.\\
This paper is organized as follows. In Section \ref{deformeddilation} we discuss the transformation of the metric components under a gauge transformations involving long-wavelength modes, for a more generic transformation than the one discussed in \cite{Matarrese:2020why} which in particular accounts for tensor modes. In Section \ref{regSVT} we show that under this deformed space dilations, the tensor perturbation of the metric tensor is unaffected and no shift is present for any finite value of the wave-number $k$. In Section \ref{Bispectrum} we discuss the transformation properties of a generic bispectrum under such a transformation. We conclude by summarizing our main results in Section \ref{conclusion}.

\section{Deformed Dilation}
\label{deformeddilation}
Let us consider  a perturbed Friedmann-Lemaitre-Robertson-Walker (FLRW) spacetime\footnote{We take the spatial $\kappa$ curvature zero for simplicity, the results could be easily extended to the case of $\kappa \neq 0$.}
\be 
ds^2=-dt^2 +a^2\,\delta_{ij}\,dx^i\,dx^j +h_{\mu\nu}\,dx^\mu\,dx^\nu \,.
\label{FLRW}
\ee
Under an infinitesimal coordinate transformation of the following type,
\begin{equation*}
x^\mu \to \tilde{x}^\mu=x^\mu+\epsilon^\mu\,,
\end{equation*}
the induced change $\Delta h=\tilde{h}(x)-h(x)$ in the metric perturbation (gauge transformation) is given by (\cite{Weinberg:2008zzc})
\begin{equation}
\begin{aligned}
\Delta h_{00}&=2\dot\epsilon^0\, ;\\
\Delta h_{0i}&=\partial_i\epsilon^0 -a^2\dot \epsilon^i\, ;\\
\Delta h_{ij}&=-2a\dot a\epsilon^0\delta_{ij} -a^2\partial_j\epsilon^i-a^2\partial_i\epsilon^j\ .
\end{aligned}
\label{hijtransfrules}
\end{equation}
Exploiting rotational invariance, it is convenient to decompose the metric perturbation into scalars, vectors and tensors (SVT) under $SO(3)$; namely, we decompose $\epsilon^\mu$  according to
\be
\epsilon^\mu=\left(\epsilon^0, \partial^i
  \epsilon+\epsilon^i_V\right) \qquad \qquad \text{where}\quad \de_i \epsilon^i_V=0 
\ee
and $h_{\mu \nu}$ as
\be
\begin{aligned}
h_{00}&=-2\phi \ ,\\
h_{0i}&=a(\partial _i F+G_i) \qquad \qquad \de_i G_i=0 \ ,\\
h_{ij}&=a^2\left (-2\, \psi\, \delta_{ij} +\partial_i\partial_j B
  +\partial_j C_i +\partial_i C_j  +D_{ij}\right ) \, \qquad \de_i
C_i=\de_j D_{ij} = \delta _{ij} D_{ij} =0 \ .
\label{so(3)decomposition}
\end{aligned}
\ee
In this way, we get the following standard transformation rules for linear perturbations (\cite{Weinberg:2008zzc})
\be 
\Delta \phi =\dot \epsilon^0 \, , \;\;\; \Delta F=\frac{1}{a}\, \epsilon^0- a \, \dot{\epsilon}\, ,\;\;\;\Delta G_i=-a\,\dot \epsilon^i_V\, ,
\ee
\begin{equation}
\Delta\psi= H\epsilon^0 \ ,\quad \Delta B=-2\epsilon \ , \quad\Delta C_i=-\epsilon^V_i \ , \quad\Delta D_{ij}=0 \, .
\label{ijgaugetransfrollo}
\end{equation}
Consider now the following transformation
\begin{equation}
\epsilon^{ i}=\lambda\, x^i+\omega^i_j \,x^j \, , \qquad \qquad
\omega^i_{i} =0 \, ,
\label{Weinbergtransformationeps}
\end{equation}
where $\lambda$ is a constant and $\omega$ a constant $3\times3$ matrix, traceless by definition\footnote{The trace part of $\omega$, one can always absorb it in $\lambda$. Indeed 
$$
\omega^i_{j}x^j=(\omega^i_{j}-\omega_k^k\delta^i_{j})x^j+\omega^k_kx^i \, ;
$$
the first term gives a traceless $\omega$ and what is left can be reabsorbed in $\lambda$.}. This can be interpreted as the leading contribution in a derivative expansion of $\epsilon^i$.
%Notice that the antisymmetric part of $\omega_{ij}$ can be eliminated by a standard rotation, while its trace corresponds to a redefinition of $\lambda$; thus we can alway take $\omega_{ij}$ to be symmetric and traceless.
Using rules (\ref{hijtransfrules}), the change of the spatial metric perturbation is given by
\begin{equation}
\Delta h_{ij}=- a^2\left (2\,\lambda
  \,\delta_{ij}+\omega^i_{j}+\omega^j_{i}\right)\,.
\label{exactklimit}
\end{equation}
Thus, only the symmetric part $\omega^{S}$ of $\omega$ contributes to the transformed  metric. Moreover, one can easily realize that the transformation (\ref{exactklimit}) can be reproduced by the following 3-parameter family of transformations of the scalar, vector and tensor parts defined in (\ref{so(3)decomposition}):
\begin{equation}
\Delta \psi=\alpha\,\lambda \ , \quad\Delta
B=\lambda \, (\alpha-1)\,x^jx^j+\gamma \, \omega_{ij}^S\, x^i\,x^j \ , \quad
\Delta C_i=(\beta-1)\,\omega_{ij}^S\,x^j \ , \quad\Delta D_{ij}=-2 \,
(\beta+\gamma) \, \omega_{ij}^S \, ,
\label{degeneracy1}
\end{equation}
with $\alpha, \,\beta, \,\gamma\in \mathbb{R}$. One should stress that the degeneracy in the above transformation rule is due the ambiguity of the decomposition (\ref{so(3)decomposition}) for the transformed metric (\ref{exactklimit}) and it is removed as soon as the coordinates transformation (\ref{Weinbergtransformationeps}) contains terms quadratic in $x^i$, or equivalently $\lambda$ and $\omega_{ij}$ becomes space-dependent (in the general case $\lambda$ and $\omega$ are functions of $x^i$). A popular choice  (\cite{ Pajer:2013ana, Dai:2015rda, Bravo:2017gct}) is to argue that a scalar perturbation $\psi_L$ and the tensor perturbation $D_L$ with a very long wavelength can always be gauged away by setting $\alpha=\beta=1$ and $\gamma=0$, thus
\be
 \Delta \psi_L=\lambda \, , \qquad \Delta D_{ij} = 2 \,
 \omega_{ij}^{S} \, , \qquad \Delta
 B=\Delta C_i =0 \, .
 \label{canc}
\ee
%
%forse bisognerebbe dire in due parole che tale scelta di $\alpha=\beta$ sembrerebbe dettata dalla rottura di simmetria. La dilatazione è rotta spontaneamente, e realizzata a livello non lineare dal dilatone che è proprio $\zeta$ e deve in quel caso shiftare.
Besides the fact that such a choice is only one among the infinitely many possible, it works only to gauge away a genuinely {\it constant} mode which is not physical\footnote{In Fourier basis this is equivalent to have a scalar or a tensor perturbation proportional to $\delta^{(3)}(\vec{k})$.}.\\

\noindent Consider now to split the scalar and tensor part of the metric perturbation in their long and short parts
\be
\psi = \psi_L + \psi_S \, , \qquad D_{ij}=
D_{ij}^{(L)}  +D_{ij}^{(S)} \, ,
\label{split1}
\ee
by using a suitable window function $W(k)=W_k$ such that
\be 
\psi_{L}(x)=\frac{1}{(2\,\pi)^3} \int d^3 k\, e^{i\, \textbf{k}\cdot\textbf{x}} \,W_k \,\psi(k) \,;
\label{split}
\ee
and similarly for the tensor part. In \cite{Pajer:2013ana} it was claimed that under a class of coordinates transformations\footnote{This class of transformations are similar to the ones used in the transition from comoving to conformal Fermi coordinates (\cite{Pajer:2013ana,Dai:2015rda}). See the Appendix.}
%({\color{red} per i segni: vedasi eq. 6.1 della tesi})
%
\be
x^i \to (1- \psi_L )\, x^i+\frac{1}{2}\,D_{j}^i{}_L\, x^j\,, 
\label{gaugetransformation}
\ee
(basically the same of (\ref{Weinbergtransformationeps}) when $\psi_L$ and $D_{ij}$ are constant) the ``long wavelength'' part in (\ref{split}) can be removed by using the transformation rules (\ref{canc}). The
problem is that
\begin{itemize}
\item
the choice that leads to (\ref{canc}) is by no means unique; for instance, by taking  $\beta=-\gamma$ and $\alpha=0$ in (\ref{degeneracy1}), then (\ref{canc}) is no longer valid: the transformations of the scalar $B$ and of the transverse vector $C_i$ reproduce (\ref{exactklimit}) with $\psi$ unchanged;
\item the cancellation can take place only in the peculiar case of purely constant $\psi_L$ and $D_{j}^i{}_L$ and this is not the case in any reasonable coordinate transformation.
\end{itemize}
As it will be shown in the next section \ref{regSVT}, when the splitting (\ref{split1}) between long and short parts is done by a physical window function, the ambiguity  (\ref{degeneracy1}) disappears and the standard transformation rules (\ref{ijgaugetransfrollo}) are recovered; thus no shift is present when a proper gradient expansion is considered. As already discussed in~\cite{Matarrese:2020why}, the transformation rules (\ref{ijgaugetransfrollo}) are precisely the ones that guarantee the gauge invariant character of scalars related to $\psi$-like fields, the comoving curvature perturbation $\zeta$ and $D_{ij}$ itself.\\
We conclude this section by underlining that the ambiguity just described was used by Weinberg to show that in the large-scale limit, under a number of technical assumptions, there is a least a conserved adiabatic mode~\cite{Weinberg:2008zzc,Weinberg:2003sw} by exploiting the residual gauge-invariance of the perturbed FLRW metric in the Newtonian gauge. But we remark that this is valid only in the exact $k=0$ limit, when $\lambda$ and $\omega$ are pure constants and so the ambiguity (\ref{degeneracy1}) is still present.

\section{Restoring the SVT Decomposition: a discontinuity in the gradient expansion}
\label{regSVT}
Let us consider a deformation of (\ref{Weinbergtransformationeps}) in the sense that now both $\lambda$ and $\omega_{ij}$ can depend on the space point $\vec{x}$, namely
\begin{equation}
\label{def_dil}
\epsilon^i= \lambda(x)\, x^i+\omega^i_{j}(x)\, x^j\,, \qquad
\omega^i_{i}=0  \, , \qquad \partial_i \omega^i_j=0\,. 
\end{equation}
To be as general as possible, we consider $\omega$ to be transverse and traceless, but not symmetric. By introducing a suitable window function $W_k$, in Fourier space $\lambda$ and  $\omega$ are written as 
\be
\lambda=\frac{1}{(2\,\pi)^3}\,\int d^3 k\,e^{i\,
  \textbf{k}\cdot\textbf{x}}\, W_k\, \lambda_k\,, \qquad 
\omega^{ij} = \frac{1}{(2\,\pi)^3}\, \int d^3 k\,e^{i\,
  \textbf{k}\cdot\textbf{x}}\, W_k\, \omega^{ij} _{\vec k} \, .
\ee
A very common choice for the window function $W_k$ is
\be
 W_k=\theta \left[\frac{1}{H}\left(k_c-k\right)\right] \ ,
\ee
where $k_c$ is a reference scale for the long-short modes splitting: modes with a wavelength smaller than $k_c>0$ are considered long and they do not contribute. However, keep in mind that the rest of this section is independent of the particular choice of $W_k$. Notice also that we have taken the function $\lambda_k$ such that it depends only on $k=|\vec{k}|$, being (for our purposes) related to the Fourier transform of $\zeta$ on super-horizon scales (\cite{Matarrese:2020why}).\\
Using the transformation (\ref{def_dil}) in (\ref{hijtransfrules}), the variation of $h_{ij}$ results in
\be
\Delta h_{ij}= -a^2 \left( \de_i \epsilon^j+  \de_j \epsilon^i
\right)= -a^2   \left[2 \, \delta_{ij} \, \lambda + 2 \, \omega_{ij}^S
  +x^i \, \de_j \lambda +x^j \, \de_i \lambda+ x^\ell\,\left(\de_i
    \omega^j_{ \ell}+ \de_j \omega^i_{\ell} \right)
\right] \, .
\label{htrans}
\ee
The Fourier transform of ${x^i}\partial \lambda_j$ entering in (\ref{htrans}), can be written as follows (\cite{Matarrese:2020why}) by using integration by parts:
\begin{equation}
x^i \, \partial_j \lambda= -\frac{1}{(2\,\pi)^3} \,\int d^3 k\,e^{i\, \vec{k}\cdot \vec{x}} \partial_{k^i} \left(k^j\, \lambda_k\right)+\text{BT}\,.
\end{equation} 
The boundary term BT is evaluated at very large $k$, where the window function vanishes: thus, terms of such a type do not contribute. Similar considerations apply to the Fourier transform of $x^\ell \de_i \omega_{j \ell}$. As a result, in Fourier space eq. (\ref{htrans}) reads 
\be
\begin{aligned}
\Delta h_{ij}(k) 
&=a^2\left[2\,\frac{k^i \, k^j}{k}\, \lambda_k'+k^i\; \partial_{k^l} \omega^{j}_l{}_{\vec k}+k^j\;\partial_{k^l}   \omega^{i}_l{}_{\vec k}\right]  \,,
\end{aligned}
\ee
where $\lambda_k' = \frac{d \lambda_k }{ d k}$. Thus, as claimed, no shift neither in $\psi_k$ nor in $D_{ij}(\vec k)$ is present and, by comparison  with eq. (\ref{hijtransfrules}), one gets the following gauge variations
\be
\Delta \psi(k)=0 \ ,\quad  \Delta B(k)=-\frac{2}{k}\,\lambda_k' \ , \quad \Delta C_i(k)=\partial_{k^l} \omega^{i}_l{}_{\vec k} \ , \quad \Delta D_{ij}(k)=0\ .
\label{gaugevariationk}
\ee
The SVT decomposition is restored and no ambiguity is present: the would-be shift of $\psi$ is actually a gradient term involving the transformation $B$, while the would-be shift of $D_{ij}$ is turned into a gradient of $\omega_{ij}$
by using 
\be
k^i\,\partial_{k^l} \omega^{i}_l{}_{\vec k}=\partial_{k^l}\left (k^i \,\omega^{i}_l{}_{\vec k}\right)=0 \,;
\ee
the first equality is obtained by considering the traceless condition $\delta_{il}\,\omega^{i}_l=0$. As a result, the shifts in (\ref{degeneracy1}) are just the artifact of the very special form (\ref{Weinbergtransformationeps}) where $\lambda$ and $\omega_{ij}$ are taken to be constant. Whenever $\lambda$ and $\omega$ acquire a space dependence, the shifts disappear and the gauge transformation cannot be used to cancel a physical long mode (that is not proportional to $\delta^{(3)}(\vec{k})$).

\section{Gauge Variation of a Correlator}
\label{Bispectrum}
We are interested in the correlation function of an operator $ {\cal O}(\vec{x}_1,...\vec{x}_N)$ built out of $\zeta$ and the tensor mode $D_{ij}$ taken as quantum field operators and evaluated by using the in-in formalism, see for instance \cite{Weinberg:2005vy}; namely
\be
{\cal O}(\vec{x}_1,...\vec{x}_N)= \zeta(\vec x_1)...\,\zeta(\vec x_M)\,D(\vec x_{M+1})... \,D(\vec x_N) \,.
\ee
In Fourier space, the case $N=3$ gives various types of bispectra.~\footnote{In Fourier space it is convenient to strip out the overall delta function according to
\be
\langle {\cal O}(\vec{k}_1,...\vec{k}_N) \rangle=(2\pi)^3 \delta^{(3)} (\vec{k}_1+...+\vec{k}_N)B_{\cal{O}}({k}_1,...,{k}_N) \ .
\nb
\ee
} The infinitesimal coordinate transformation (\ref{def_dil}) can be generalized at the non-linear level by
\be
\label{Nonlinear_dil}
\tilde x^i= e^{\lambda}\, x^i+(1-e^{\omega})|_{ij}\, x^j\,, \qquad g_{ij}=a^2 \, e^{2 \, \zeta}\,\delta_{ij}+h_{ij}+\frac{1}{2}\,h_{il}\,h_{lj}+\frac{1}{6}\,h_{il}\,h_{lm}\,h_{mj}\, ,
\ee
with $\zeta$, $\lambda$ and $\omega$ dependent on the spacetime point. Such a transformation represents the non-linear extension of  the linear deformed dilatation used to connect the comoving gauge\footnote{In our convention the comoving gauge is defined as the one in which $B$ and the peculiar velocity are set to zero.} with CFC-like reference frame.
Let us consider the gauge variation of a correlator as the the difference of the expectation value of the operator $ {\cal O'}(\vec{x}_1,...\vec{x}_N)$ in the new coordinates defined by (\ref{Nonlinear_dil})  and the original operator $ {\cal O}(\vec{x}_1,...\vec{x}_N)$. % In this section we show that in the new frame, at quadratic order in perturbation theory, $\psi(/D)$ equations of motion are the same expected in the old frame. If this last check will be satisfied, automatically the gauge variation
\be
\Delta O_\text{gauge}= \left\langle {\cal O}'(\vec{x}_1,...\vec{x}_N)\right\rangle -\left\langle {\cal O}(\vec{x}_1,...\vec{x}_N)\right\rangle \, .
\label{correlatorshift}
\ee  
%
% is identically zero and so is the bispectrum
% For our purposes, the operator ${\cal O}$ can be considered a general combination of operators $\zeta$ and $D$:
The action describing gravity and the inflaton field is invariant under  a coordinate transformation and, up to a boundary term,  it can be written in the ADM form as~\cite{PhysRev.116.1322,Maldacena:2002vr}
\be
\label{Action}
S=\int \, d^4x \, \sqrt{h} \, N\,\left[ R^{(3)}+K_{ij} \,
  K^{ij}-K^2+{\cal L}_m\right] \equiv \int \, d^4x \, \sqrt{h(x)} \,
{\cal S}(x)\,,
\ee
where $h$ is the spatial metric determinant and $K_{ij}$ is the extrinsic curvature tensor of the hypersurface (see footnote \ref{footnote1}) of equation $t=\text{constant}$, while ${\cal L}_m$ is the Lagrangian for the inflaton field $\phi$. The 3-scalar ${\cal S}$ in  (\ref{Action}) can be expanded as 
\be
\label{tr_prop}
\tilde {\cal S}(\tilde x) \equiv {\cal S}(x)= \bar {\cal S}(t)+{\cal S}^{(1)}(x)+{\cal S}^{(2)}(x)+\dots\,,
\ee
where $\bar {\cal S}(t)$ contains only background quantities, while ${\cal S}^{(n)}(x)$ is of order $n$ in perturbations. It is convenient to define the following gauge variation
\be 
\Delta_{\cal S}=\sqrt{\tilde h (x)}\, \tilde {\cal S}(x)-\sqrt{h(x)}\,\, {\cal S}(x)\,,
\ee
which gives for the change of the action $\Delta_{\text{action}}=\int \, d^4x \, \Delta_{\cal S}$. Splitting also $\Delta_{S}$ into background and $n$-th order perturbations as done for $\cal S$ in (\ref{tr_prop}), the change of the spatial coordinates induces the following variation $\Delta_{\cal S}$ up to third order
\be
\label{dletaS_res_back_1_2}
\begin{cases}
&\bar \Delta_{\cal S}=0\,,\\
& \\
&\Delta_{\cal S}^{(1)}=
a^3 \,\partial_i \left(\bar{{\cal S}}\, \lambda\, x^i \right)\,,\\
& \\
&\Delta_{\cal S}^{(2)}= -a^3 \, \partial_i \left[\left({\cal S}^{(1)}+3\,\bar{{\cal S}}\zeta \right)\,\left(\lambda\, x^i+\omega^i_{j}\,x^j \right)\right]\,,\\
& \\
& \Delta_{\cal S}^{(3)}=- a^3 \, \partial_i \left[\left(\frac{9}{2}\, \bar{{\cal S}}\, \zeta^2+3\, {\cal S}^{(1)}\,\zeta+{\cal S}^{(2)} \right)\,\left(\lambda\, x^i+\omega^i_{j}\,x^j \right)\right]\, ,
\end{cases}
\ee
where for simplicity we have omitted\footnote{Being $\lambda$-$\omega$ defined by long modes only, $\lambda^n$-$\omega^n$ ($n>1$) vertices should imply correlators with two and three squeezed momenta that are not relevant for the consistency relation.} all the quadratic and cubic terms in $\lambda$-$\omega$. The botton line is that  all the new terms in the cubic action introduced by the deformed dilation can be written as total spatial derivatives. As shown in~\cite{Matarrese:2020why} the gauge variation $\Delta O_\text{gauge}$ can be written as the commutator of ${\cal O}(\vec{x}_1,...\vec{x}_N)$ with $\Delta_\text{action}$ which vanishes being $\Delta_{\cal S}$ a total spatial derivative.
% and as such they cannot contribute to the non-linear equations of motion. {\color{red} The quantum expectation value is computed by using the in-in formalism,
%the gauge variation will be automatically zero~\cite{Matarrese:2020why}, namely}
%\be
%\left\langle {\cal O}'(\vec{x}_1,...\vec{x}_N)\right\rangle -\left\langle {\cal O}(\vec{x}_1,...\vec{x}_N)\right\rangle \equiv 0 \ ,
%\ee
%as stated above.\\
%\noindent Given a gauge transformation acting on a field $\cal O$ as
%\be
%{\cal O}'(x)={\cal O}(x)-\xi^j \, \partial_j  {\cal O}(x) \,,
%\ee
%
%one could imagine that this could alter the shape of  squeezed bispectrum. However, in \cite{Matarrese:2020why} it is explicitly shown that under an infinitesimal transformation of type $\epsilon^i=\lambda \,x^i$, thanks to the fact that $\zeta$ is a scalar, the 3-point function shifts with a boundary term in Fourier space. The same can be easily generalized to the case of $\epsilon^i=\lambda\, x^i+\omega_{j}^i\, x^j$.
In conclusion, one cannot gauge away the long modes of the fields at least at first order in perturbation theory, so the squeezed bispectrum cannot be canceled.

\section{Conclusion}
\label{conclusion}
Measuring the primordial non-Gaussianity remains one of the most important goals to study the physics of the Early Universe. In single-field inflationary model the squeezed limit is completely fixed in a model independent way thanks to the consistency relation, but its physical observability has been criticized, limiting the contributions to observed correlations to projection effects such as gravitational lensing and redshift perturbations (\cite{Pajer:2013ana}). As discussed in \cite{Matarrese:2020why}, in this debate it is crucial to determine how a very long perturbation affects the quantities of physical interest. In this paper we have analysed carefully the transformation properties of cosmological observables such as the curvature perturbation $\zeta$, the tensor perturbation $D$ and their correlation functions, thereby generalizing the results of \cite{Matarrese:2020why} where the analysis was done for correlators involving only $\zeta$'s. In this case the infinitesimal diffeomorphism is generalized by equation (\ref{Weinbergtransformationeps}) and, in the same way, the result is that no shift both in $\zeta$ and in $D$ is found, independently of the filter used to select long modes, excluding the case of infinitely long-wavelength (hence non-physical) perturbations. The latter is the main ingredient often stated for canceling tensor-scalar $f_{\text{NL}}$, but we have seen that this is not consistent with a CFC-like transformation. We think that the problem is
%this result comes from a misleading interpretation of the Weinberg transformation and 
the role played by a  constant spatial dilatation in single-field inflation\footnote{It is well-known that single-field cubic interactions are conformally symmetric. In \cite{Hinterbichler:2013dpa}, it is shown how to extract the scalar consistency relations by using dilatation symmetry itself and the related Ward identities. }.\\
The transformation rules for the SVT elements in Fourier space presented in \cite{Matarrese:2020why} have been generalized in eq. (\ref{gaugevariationk}), showing once again that no shift is present neither in $\psi$ nor in $D$. It has also been shown explicitly that the equations of motion are not affected by a gauge transformation of type (\ref{Weinbergtransformationeps}), implying that the correlator is unaffected according to eq. (\ref{correlatorshift}). Indeed, to cancel the bispectrum it has been used that this difference gives $-\left\langle {\cal O}(\vec{x}_1,...\vec{x}_N)\right\rangle$ (\cite{Tanaka_2011, Pajer:2013ana, Dai:2015rda, Bravo:2017gct}), but we have shown this to be not the case.\\
The outcome of our study is that all the squeezed bispectra, involving both $\zeta$ and $D$, cannot be gauged away and they remain physical observables, analogously to what obtained in \cite{Matarrese:2020why} for $B_{\zeta\zeta\zeta}$. This has a remarkable impact on future observations of a primordial gravitational-wave background. Consistency relations remain a very important tool to study Early Universe physics.\\

\textbf{Acknowledgements}: N.B. and S.M. acknowledge support from the COSMOS network\\ (www.cosmosnet.it) through the ASI (Italian Space Agency) Grants 2016-24-H.0, 2016-24-H.1-2018 and 2019-9-HH.0.

\appendix 
\section{Appendix: CFC transformation}
\noindent In this appendix we give the structure of $\lambda$ and $\omega$ related to the CFC expansion. At first order of the CFC transformation, $\lambda$ and $\omega$ reduce to be exactly $\zeta$ and $\frac{D_{ij}}{2}$, as in eq. (\ref{gaugetransformation}). However, the first-order analysis gives rise to two main issues:
\begin{enumerate}
\item as just demonstrated, we get a discontinuity in the gradient expansion;
\item the typical structure of the CFC metric ($g_{ij}^F \sim O(x_F^2)$) is not reproduced. 
\end{enumerate}
For this reason, we are forced to consider the transformation to the CFC frame up %to first order in scalar and tensor perturbations and up 
to third order in the CFC series. The scalar part has already been discussed in \cite{Matarrese:2020why}, so we can consider only the tensor perturbation part of the transformation:
\be
\begin{aligned}
\Delta x^k_F=&\Delta x^k+\frac{1}{2} D^k_i\bar\Delta x^i\bigg|_p +\frac{1}{4}\bar\Delta x^i \bar\Delta x^ j  ( \partial_i D^k_{j}+\partial_j D^k_{i}-\partial^k D_{ij} ) \bigg|_p+\\
&+\frac{1}{12}\bar\Delta x^i \bar\Delta x^ j \bar\Delta x^ l  ( \partial_l\partial_i D^k_{j}+\partial_l\partial_j D^k_{i}-\partial_l\partial^k D_{ij} ) \bigg|_p  \ ,
\end{aligned}
\label{tensorCFC}
\ee
where $\Delta x=x(\tau)-p(\tau)$ is the deviation from a central world-line and $\bar\Delta$ is its background value. The transformation (\ref{tensorCFC}) can be SVT decomposed in Fourier space as follows,
\be 
\epsilon_k=-\frac{1}{12}\,\frac{1}{k^2}\, \sum_{s=\pm2}  \left(D_k^{(s)}-k \, D_{k}^{(s)}{}'\right)\,,
\ee
where we considered $D_{ij}{}_{\vec{k}}=\sum_{s=\pm2} \varepsilon^{(s)}_{ij}\,D_k^{(s)}$ in the standard convention for spin-2 polarization tensors, $\varepsilon^{(s)}_{ij} \,\varepsilon^{(s')\,*}_{ij}= 2 \, \delta^{ss'}$ and 
\be
\epsilon^i_V=-\frac{i}{12}\left (10 \partial_{k_l} D_l^i+2k_m \partial_{k_m}\partial_{k_j}D_{ij}\right) \ .
\label{epsilonVk}
\ee
This allows the extension of the results presented in \cite{Matarrese:2020why}, where only the scalar sector was considered. Notice that, since $\partial_i \epsilon^i_V=0$,  (\ref{epsilonVk}) can be always rewritten as $\epsilon^i_V=A^i_k x^k$ with $A$ depending on the space-time point but transverse and traceless, so playing the role of $\omega$ in (\ref{def_dil}). Thus, to reproduce the proper local structure of the metric tensor $g_{ij}^F \sim O(x_F^2)$ we find an interesting mixing: the tensor degrees of freedom start to influence both the scalar and vector sectors.

\printbibliography

\end{document}